\documentclass{IeiceModify}
\usepackage[dvips]{graphicx,color}
\usepackage{latexsym}
\usepackage[dvips]{graphicx,color}
\usepackage{latexsym}
\field{A}
\title[On Two Strong Converse Theorems 
for Discrete Memoryless Channels]
{On Two Strong Converse Theorems 
for Discrete Memoryless Channels} 
\titlenote{
This work was presented in part at the 28rd Symposium on 
Information Theory and Its Applications, Onna, Okinawa, Japan, 
Nov. 20-23, 2005.
}
\authorlist{%
\authorentry{Yasutada OOHAMA}{%m
}{labelA}
}
\breakauthorline{3}
\affiliate[labelA]
{
The author is with 
The University of Electro-Communications,
1-5-1 Chofugaoka, Chofu-Shi, Tokyo 182-8585, Japan
}

\begin{document}
\newcommand{\bc}{\begin{center}}  %
\newcommand{\ec}{\end{center}}
\newcommand{\befi}{\begin{figure}[h]}  %
\newcommand{\enfi}{\end{figure}}
\newcommand{\bsb}{\begin{shadebox}\begin{center}}   %
\newcommand{\esb}{\end{center}\end{shadebox}}
\newcommand{\bs}{\begin{screen}}     %
\newcommand{\es}{\end{screen}}
\newcommand{\bib}{\begin{itembox}}   %
\newcommand{\eib}{\end{itembox}}
\newcommand{\bit}{\begin{itemize}}   %
\newcommand{\eit}{\end{itemize}}
\newcommand{\defeq}{\stackrel{\triangle}{=}}
\newcommand{\qed}{\hbox{\rule[-2pt]{3pt}{6pt}}}
\newcommand{\beq}{\begin{equation}}
\newcommand{\eeq}{\end{equation}}
\newcommand{\beqa}{\begin{eqnarray}}
\newcommand{\eeqa}{\end{eqnarray}}
\newcommand{\beqno}{\begin{eqnarray*}}
\newcommand{\eeqno}{\end{eqnarray*}}
\newcommand{\ba}{\begin{array}}
\newcommand{\ea}{\end{array}}
\newcommand{\vc}[1]{\mbox{\boldmath $#1$}}
\newcommand{\lvc}[1]{\mbox{\footnotesize\boldmath $#1$}}

\newcommand{\wh}{\widehat}
\newcommand{\wt}{\widetilde}
\newcommand{\ts}{\textstyle}
\newcommand{\ds}{\displaystyle}
\newcommand{\scs}{\scriptstyle}
\newcommand{\vep}{\varepsilon}
\newcommand{\rhp}{\rightharpoonup}
\newcommand{\cl}{\circ\!\!\!\!\!-}
\newcommand{\bcs}{\dot{\,}.\dot{\,}}
\newcommand{\eqv}{\Leftrightarrow}
\newcommand{\leqv}{\Longleftrightarrow}
\newtheorem{co}{Corollary} 
\newtheorem{lm}{Lemma} 
\newtheorem{Ex}{Example} 
\newtheorem{Th}{Theorem}
\newtheorem{df}{Definition} 
\newtheorem{pr}{Property} 
\newtheorem{rem}{Remark} 

\newcommand{\lcv}{convex } 

\newcommand{\hugel}{{\arraycolsep 0mm
                    \left\{\ba{l}{\,}\\{\,}\ea\right.\!\!}}
\newcommand{\Hugel}{{\arraycolsep 0mm
                    \left\{\ba{l}{\,}\\{\,}\\{\,}\ea\right.\!\!}}
\newcommand{\HUgel}{{\arraycolsep 0mm
                    \left\{\ba{l}{\,}\\{\,}\\{\,}\vspace{-1mm}
                    \\{\,}\ea\right.\!\!}}
\newcommand{\huger}{{\arraycolsep 0mm
                    \left.\ba{l}{\,}\\{\,}\ea\!\!\right\}}}
\newcommand{\Huger}{{\arraycolsep 0mm
                    \left.\ba{l}{\,}\\{\,}\\{\,}\ea\!\!\right\}}}
\newcommand{\HUger}{{\arraycolsep 0mm
                    \left.\ba{l}{\,}\\{\,}\\{\,}\vspace{-1mm}
                    \\{\,}\ea\!\!\right\}}}

\newcommand{\hugebl}{{\arraycolsep 0mm
                    \left[\ba{l}{\,}\\{\,}\ea\right.\!\!}}
\newcommand{\Hugebl}{{\arraycolsep 0mm
                    \left[\ba{l}{\,}\\{\,}\\{\,}\ea\right.\!\!}}
\newcommand{\HUgebl}{{\arraycolsep 0mm
                    \left[\ba{l}{\,}\\{\,}\\{\,}\vspace{-1mm}
                    \\{\,}\ea\right.\!\!}}
\newcommand{\hugebr}{{\arraycolsep 0mm
                    \left.\ba{l}{\,}\\{\,}\ea\!\!\right]}}
\newcommand{\Hugebr}{{\arraycolsep 0mm
                    \left.\ba{l}{\,}\\{\,}\\{\,}\ea\!\!\right]}}
\newcommand{\HUgebr}{{\arraycolsep 0mm
                    \left.\ba{l}{\,}\\{\,}\\{\,}\vspace{-1mm}
                    \\{\,}\ea\!\!\right]}}

\newcommand{\hugecl}{{\arraycolsep 0mm
                    \left(\ba{l}{\,}\\{\,}\ea\right.\!\!}}
\newcommand{\Hugecl}{{\arraycolsep 0mm
                    \left(\ba{l}{\,}\\{\,}\\{\,}\ea\right.\!\!}}
\newcommand{\hugecr}{{\arraycolsep 0mm
                    \left.\ba{l}{\,}\\{\,}\ea\!\!\right)}}
\newcommand{\Hugecr}{{\arraycolsep 0mm
                    \left.\ba{l}{\,}\\{\,}\\{\,}\ea\!\!\right)}}

\newcommand{\hugepl}{{\arraycolsep 0mm
                    \left|\ba{l}{\,}\\{\,}\ea\right.\!\!}}
\newcommand{\Hugepl}{{\arraycolsep 0mm
                    \left|\ba{l}{\,}\\{\,}\\{\,}\ea\right.\!\!}}
\newcommand{\hugepr}{{\arraycolsep 0mm
                    \left.\ba{l}{\,}\\{\,}\ea\!\!\right|}}
\newcommand{\Hugepr}{{\arraycolsep 0mm
                    \left.\ba{l}{\,}\\{\,}\\{\,}\ea\!\!\right|}}

\newenvironment{jenumerate}
	{\begin{enumerate}\itemsep=-0.25em \parindent=1zw}{\end{enumerate}}
\newenvironment{jdescription}
	{\begin{description}\itemsep=-0.25em \parindent=1zw}{\end{description}}
\newenvironment{jitemize}
	{\begin{itemize}\itemsep=-0.25em \parindent=1zw}{\end{itemize}}
\renewcommand{\labelitemii}{$\cdot$}

\newcommand{\iro}[2]{{\color[named]{#1}#2\normalcolor}}
\newcommand{\irr}[1]{{\color[named]{Red}#1\normalcolor}}
\newcommand{\irg}[1]{{\color[named]{Green}#1\normalcolor}}
\newcommand{\irb}[1]{{\color[named]{Blue}#1\normalcolor}}
\newcommand{\irBl}[1]{{\color[named]{Black}#1\normalcolor}}

\newcommand{\irp}[1]{{\color[named]{Yellow}#1\normalcolor}}
\newcommand{\irO}[1]{{\color[named]{Orange}#1\normalcolor}}
\newcommand{\irBr}[1]{{\color[named]{Purple}#1\normalcolor}}
\newcommand{\IrBr}[1]{{\color[named]{Purple}#1\normalcolor}}
%
%
%
%
%-----------------indention environment-------------------%
\newenvironment{indention}[1]{\par
\addtolength{\leftskip}{#1}\begingroup}{\endgroup\par}
%form: \begin{indention}{2.3cm}
%      \end{indention}
%---------------------------------------------------------%
%
%----------------- namelist environment-------------------%
\newcommand{\namelistlabel}[1]{\mbox{#1}\hfill} 
\newenvironment{namelist}[1]{%
\begin{list}{}
{\let\makelabel\namelistlabel
\settowidth{\labelwidth}{#1}
\setlength{\leftmargin}{1.1\labelwidth}}
}{%
\end{list}}
%form: \begin{namelist}{width}
%---------------------------------------------------------%
%
%\def\BibTeX{{\rm B\kern-.05em{\sc i\kern-.025em b}\kern-.08em
%    T\kern-.1667em\lower.7ex\hbox{E}\kern-.125emX}}
%
%\newcommand{\bfig}{\begin{figure}[f]}
%\newcommand{\efig}{\end{figure}}
%\setcounter{page}{-41}
\newcommand{\bfig}{\begin{figure}[t]}
\newcommand{\efig}{\end{figure}}
\setcounter{page}{1}

\newtheorem{theorem}{Theorem}

\newcommand{\ep}{\mbox{\rm e}}

\newcommand{\Exp}{\exp%_2
}

\newcommand{\idenc}{{\varphi}_n}
\newcommand{\resenc}{%\tilde
{\varphi}_n}
\newcommand{\ID}{\mbox{\scriptsize ID}}
\newcommand{\TR}{\mbox{\scriptsize TR}}
\newcommand{\Av}{\mbox{\sf E}}

\newcommand{\Vl}{|}
\newcommand{\Ag}{(R,P_{X^n}|W^n)}
\newcommand{\Agv}[1]{({#1},P_{X^n}|W^n)}
\newcommand{\Avw}[1]{({#1}|W^n)}

\newcommand{\Jd}{X^nY^n}
\newcommand{\IdR}{r_n}

\newcommand{\Index}{{n,i}}

\newcommand{\cid}{C_{\mbox{\scriptsize ID}}}
\newcommand{\cida}{C_{\mbox{{\scriptsize ID,a}}}}

\newcommand{\SP}{\mbox{{\scriptsize sp}}}
\newcommand{\mSP}{\mbox{{\scriptsize sp}}}

\def\de{\stackrel{\rm def}{=}}
\def\no{\nonumber}
\def\A{{\cal A}}
\def\B{{\cal B}}
\def\E{{\cal E}}
\def\F{{\cal F}}
\def\Q{{\cal Q}}
\def\X{{\cal X}}
\def\Y{{\cal Y}}
\def\Z{{\cal Z}}
\def\C{{\cal C}}
\def\K{{\cal K}}
\def\L{{\cal L}}
\def\M{{\cal M}}
\def\U{{\cal U}}
\def\R{{\cal R}}
\def\S{{\cal S}}
\def\T{{\cal T}}
\def\D{{\cal D}}
\def\Dn{{\cal D}_n}
\def\B{{\cal B}}
\def\BW{{\mbox{\bf W}}}
\def\BM{{\mbox{\bf M}}}
\def\BA{{\mbox{\bf A}}}
\def\BB{{\mbox{\bf B}}}
\def\BE{{\mbox{\bf E}}}
\def\BZ{{\mbox{\bf Z}}}
\def\BW{{\mbox{\bf W}}}
\def\BV{{\mbox{\bf V}}}
\def\BR{{\mbox{\bf R}}}
\def\BU{{\mbox{\bf U}}}
\def\BX{{\mbox{\bf X}}}
\def\BY{{\mbox{\bf Y}}}
\def\bx{{\mbox{\bf x}}}
\def\bm{{\mbox{\bf m}}}
\def\bc{{\mbox{\bf c}}}
\def\Ba{{\mbox{\bf a}}}
\def\tba{{\mbox{\tiny{\bf a}}}}
\def\bb{{\mbox{\bf b}}}
\def\tbb{{\mbox{\tiny{\bf b}}}}
\def\by{{\mbox{\bf y}}}
\def\bz{{\mbox{\bf z}}}
\def\bw{{\mbox{\bf w}}}
\def\bv{{\mbox{\bf v}}}
\def\bu{{\mbox{\bf u}}}

\arraycolsep 0.5mm

\maketitle

\begin{summary}
In 1973, Arimoto proved the strong converse theorem for the discrete 
memoryless channels stating that when transmission rate $R$ is above 
channel capacity $C$, the error probability of decoding goes to one as 
the block length $n$ of code word tends to infinity. He proved the 
theorem by deriving the exponent function of error probability of 
correct decoding that is positive if and only if $R>C$. Subsequently, in 
1979, Dueck and K\"orner determined the optimal exponent of correct 
decoding. Arimoto's bound has been said to be equal to the bound of 
Dueck and K\"orner. However its rigorous proof has not been presented 
so far. In this paper we give a rigorous proof 
of the equivalence of Arimoto's bound to that of Dueck and K\"orner.
\end{summary}
\begin{keywords} 
%Identification via channels, 
Strong converse theorem, discrete memoryless channels, exponent 
of correct decoding
%strong converse theorem, error exponent 
\end{keywords}

\section{Introduction}

%In some the data transmission using noisy channels, when transmission rate 
%is above capacity, the error probability of decoding goes to one as 
%the block length of code words tend to infinity. 

In some class of noisy channels the error probability of decoding goes 
to one as the block length $n$ of transmitted codes tends to infinity at 
rates above the channel capacity. This is well known as a strong converse 
theorem for noisy channels. In 1957, Wolfowitz \cite{ww} proved the strong 
converse theorem for discrete of memoryless channels(DMCs). His result 
is the first result on the strong converse theorem.

In 1973, Arimoto \cite{ari} obtained some stronger 
result on the strong converse 
theorem for DMCs. He proved that the error probability of decoding goes 
to one exponentially and derived a lower bound of the exponent function. 
To prove the above strong converse theorem he introduced an interesting 
bounding technique based on a symmetrical structure of the set of 
transmission codes. Using this bounding method and an analytical 
argument on convex functions developed by Gallager \cite{ga}, 
he derived the lower bound.

Subsequently, Dueck and K\"orner \cite{dk} determined the optimal 
exponent function for the error probability of decoding to go to one. 
They derived the result by using a combinatorial method base on the type 
of sequences. Their method is quite different from the method of Arimoto 
\cite{ari}. In their paper, Dueck and K\"orner \cite{dk} stated that 
their optimal bound can be proved to be equal to the lower bound of 
Arimoto \cite{ari} by analytical computation. However, after their 
statement we have found no rigorous proof of the above equality so far 
in the literature.

In this paper we give a rigorous proof of the equality of the 
lower bound of Arimoto \cite{ari} to that of the optimal bound of Dueck and 
K\"orner \cite{dk}. To prove the above equality, we need to prove the convex 
property of the optimal exponent function. We prove this by an {\it 
operational meaning} of the optimal exponent function. Contrary to their 
statement, our arguments of the proof are {\it not completely analytical}. 
A dual equivalence of two exponent functions was established by Csisz\'ar 
and K\"orner \cite{ck} on the exponent functions for the error 
probability of decoding to go to zero at rates below capacity. Their 
arguments of the proof of equivalence are {\it completely analytical}. 
We compare our arguments to their ones to clarify an essential 
difference between them.  

%The above method we adopt for the proof of the convexity implies 
%that 
%we use for the 
%proof of the convexity of the optimal exponent function 
%\section{Coding of DiscreteMCs}

\section{Coding Theorems for Discrete Memoryless Channels}
We consider the discrete memoryless channel with the input set 
${\cal X}$ and the output set ${\cal Y}$. We assume that 
${\cal X}$ and ${\cal Y}$ are finite sets. 
Let $X^n$ be a random variable taking values in ${\cal X}^n$.
Suppose that $X^n$ has a probability distribution on ${\cal X}^n$ 
denoted by 
$P_{X^n}=$ 
$\left\{P_{X^n}({\vc x}) 
 \right\}_{{\lvc x} \in {\cal X}^n}$.
Let $Y^n\in {\cal Y}^n$  be a random variable 
obtained as the channel output by connecting 
$X^n$ to the input of channel. 
We write a conditional distribution of $Y^n$ on given $X^n$ 
as 
$W^n=$ 
$\left\{W^n({\vc y}| {\vc x})
 \right\}_{({\lvc x},{\lvc y})\in {\cal X}^n\times{\cal Y}^n}$.
A noisy channel is defined by a sequence of stochastic matrices 
$\left\{W^n \right\}_{n=1}^{\infty}$. In particular, 
a stationary discrete memoryless channel is defined by 
a stochastic matrix with input set ${\cal X}$ and output set
${\cal Y}$. We write this stochastic matrix as 
$W=$$\left\{W(y|x) \right\}_{(x,y)\in {\cal X}^n\times{\cal Y}^n }$.

Information transmission using the above noisy channel is formulated as 
follows. Let $\M_{n}$ be a message set to be transmitted through the 
channel. Set $M_n=|{\cal M}_n|$. For given $W$, 
a $(n,M_n, \varepsilon_{n})$-code is a set of $\{({\vc x}(m),$ 
   $\D({m})$, 
   $m \in {\cal M}_n, \}$ 
that satisfies the following:
$$
\ba{ll}
\mbox{1)}& {\vc x}(m) \in {\cal X}^n\,, 
\vspace{1mm}\\    
\mbox{2)}& \D(m), m \in {\cal M}_n 
           \mbox{ are disjoint subsets of }{\cal Y}^n,   
\vspace{1mm}\\    
\mbox{3)}& \ds \varepsilon_{n}=\frac{1}{M_n}\sum_{ m\in {\cal M}_n}
           W^n(({\D}(m))^c |{\vc x} (m) )\,,
\ea
$$
where $\D(m), m \in {\cal M}_n $ are decoding regions of the code and  
$\varepsilon_n$ is the error probability of decoding.

A transmission rate $R$ is achievable if 
there exists a sequence of $(n,M_n,$ $\varepsilon_{n})$-codes, 
$n=1,$ $2,\cdots$ such that 
\beq
\ds\limsup_{n\to\infty}\varepsilon_{n} =0\,,%\leq \varepsilon\,,
\ds\liminf_{n\to\infty} \frac{1}{n} \log M_n \geq R\,.
\eeq
Let the supremum of achievable transmission rate $R$ be denoted by 
${C}(W)$, which we call the channel capacity. It is well known that   
${C}(W)$ is given by the following formula: 
\beq
C(W)=\max_{P\in {\cal P}({\cal X})}I(P,W)\,, 
\eeq
where ${\cal P}({\cal X})$ is a set of probability distribution 
on ${\cal X}$ and $I(P,W)$ stands for a mutual information between 
$X$ and $Y$ when input distribution of $X$ is $P$.

To examine an asymptotic behavior of $\varepsilon_n$ for 
large $n$ at $R<C(W)$, we define the following quantities.
For give $R\geq 0$, the quantity $E$ is achievable error exponent 
if there exits a sequence of $(n,M_n,$ $\varepsilon_{n})$-codes,
$n=1,2,\cdots$ such that
$$
\ds\liminf_{n\to\infty} \frac{1}{n} \log M_n \geq R\,,
\: 
\ds\liminf_{n\to\infty}
\left(-\frac{1}{n}\right)\log\varepsilon_{n} \geq E\,. 
$$
The supremum of the achievable error exponent $E$ is denoted by 
${E}^{*}(R|W)$. Several lower and upper bounds of   
${E}^{*}(R|W)$ have been derived so far. An explicit form of 
${E}^{*}(R|W)$ is known for large $R$ below $C(W)$. 
An explicit formula of ${E}^{*}(R|W)$ for all $R$ below $C(W)$ 
has been unknown yet.

\section{Strong Converse Theorems for Discrete Memoryless Channels}

Wolfowitz \cite{ww} first established the strong converse theorem for 
DMCs by proving that when $R>C(W)$, we have 
$\lim_{n\to\infty}\varepsilon_n=1$. 
When strong converse theorem holds, we are interested 
in a rate of convergence for the error probability 
of decoding to tend to one as $n \to \infty$ for$R>C(W)$. 
To examine the above rate of convergence, we define the 
following quantity. 
For give $R\geq 0$, the quantity $G$ is achievable exponent 
if there exits a sequence of 
$(n,M_n,$ $\varepsilon_{n})$-codes,$n=1,$ $2,\cdots$ 
such that
$$
\ds\liminf_{n\to\infty} \frac{1}{n} \log M_n \geq R \,, 
\ds\limsup_{n\to\infty}
\left(-\frac{1}{n}\right)\log(1-\varepsilon_{n}) \leq G\,. 
$$
The infmum of the achievable exponent $G$ is denoted by 
${G}^{*}(R|W)$. This quantity has the following property.
\begin{pr}\label{pr:pr0}{
The function ${G}^{*}(R|W)$ is a monotone increasing and convex
function of $R$.
}
\end{pr}

{\it Proof:} By definition it is obvious that ${G}^{*}(R|W)$ 
is a monotone 
increasing function of $R$. To prove the convexity 
fix two positive rates $R_1,R_2$ arbitrary.
For each $R_i,i=1,2$, we consider the infimum of the achievable 
exponent function ${G}^{*}(R_i|W)$. 
By the definitions of ${G}^{*}(R_i|W),i=1,2$, 
for each $i=1,2$, there exists 
a sequence of $(n,M_n^{(i)},$ $\varepsilon_{n}^{(i)})$-codes, 
$n=1,2,\cdots,$ such that  
\beqno
\ds\liminf_{n\to\infty} \frac{1}{n} \log M_n^{(i)} 
&\geq& R_i\,, 
\nonumber\\
\ds\limsup_{n\to\infty}
\left(-\frac{1}{n}\right)
\log\left(1-\varepsilon_{n}^{(i)}\right) 
&\leq& G^{\ast}(R_i|W)\,. 
\eeqno
Fix any $\lambda_i, i=1,2$ with $\lambda_1+\lambda_2=1$
and set $n_i=\lfloor \lambda_i n \rfloor$, where 
$\lfloor a \rfloor$ stands for the integer part of $a$.
Set $\nu=n-n_1-n_2$. It is obvious that $\nu\in\{0,1,2\}$.

Next, we consider the code obtained by concatenating 
$(n_i,M_n^{(i)},$ $\varepsilon_{n}^{(i)})$-codes for $i=1,2$.
If $\nu=1$ or 2, we further append %a proper 
$(\nu,1,0)$-code. For the above constructed 
$(n,M_n,$ $\varepsilon_{n})$-code we have 
\beqno
M_{n} = \prod_{i=1,2}M_{n_i}^{(i)}, \quad 
1-\varepsilon_n = \prod_{i=1,2}
\left(1-\varepsilon_{n_i}^{(i)}\right).
\eeqno
Then, we have 
\beqno
& &\ds\liminf_{n\to\infty} \frac{1}{n} \log M_{n}
\nonumber\\
&= & \sum_{i=1,2}\ds\liminf_{n\to\infty} 
\frac{n_i}{n}\cdot\frac{1}{n_i} \log M_{n_i}^{(i)} 
%\nonumber\\
\geq  \sum_{i=1,2}\lambda_i R_i\,,
\nonumber\\
& &\ds\limsup_{n\to\infty}
\left(-\frac{1}{n}\right)\log\left(1-\varepsilon_{n}\right)
\nonumber\\
&=&\sum_{i=1,2}
\limsup_{n\to\infty}
\frac{n_i}{n}\cdot \left(-\frac{1}{{n_i}}\right)
\log\left(1-\varepsilon_{n_i}^{(i)}\right)
\nonumber\\
&\leq& \sum_{i=1,2}\lambda_i G^{*}(R_i|W)\,.
\eeqno
Hence, we have 
$$
\sum_{i=1,2}\lambda_i G^{*}(R_i|W)\geq  
G^{*}\left(\sum_{i=1,2}\lambda_iR_i \Hugepl W\right)\,,
$$
which implies the convexity of $G^{*}(R_i|W)$. 
\hfill\QED

Arimoto \cite{ari} derived a lower bond of ${G}^{*}(R|W)$. 
To state his result we define 
some functions. For $\delta\in [-1,+\infty)$, define   
\newcommand{\prmt}{\delta}
\beqno
&&{J}_{\prmt}(P\Vl W) 
 \defeq -\log%\left\{
\sum_{y \in {\cal Y} }
\left[\sum_{x \in {\cal X}} 
P(x)W(y|x)^{\frac{1}{1+\prmt}} \right]^{1+\prmt}
%\right\}
\,,
\\
& &{\empty}{F}_{\prmt}(R,P\Vl W)\defeq \prmt R + {J}_{\prmt}(P\Vl W),
\\
& &{\empty}{G}_{\prmt}(R\Vl W) \defeq \min_{P\in {\cal P}({\cal X})}
{\empty}{F}_\prmt(R,P\Vl W)\,.
\eeqno
Furthermore, set
\beqno
{\empty}{G}(R\Vl W)
&\defeq& 
\max_{-1\leq \prmt \leq 0}
{\empty}{G}_{\prmt}(R\Vl W)
\\
&=& 
\max_{-1\leq \prmt \leq 0}
\min_{P\in {\cal P}({\cal X})}{\empty}{F}_{\prmt}(R,P\Vl W)
\\
&=&
\max_{-1\leq \prmt \leq 0}
\left[
{-\prmt R}+\min_{P\in{\cal P}({\cal X})}
{J}_{\prmt}(P\Vl W)
\right]\,.
\eeqno

According to Arimoto \cite{ari}, the following property holds. 
\begin{pr}{\label{pr:pr4z}} 
The function ${G}(R\Vl W)$ is a monotone increasing and convex 
function of $R$ and is positive if and only if $R > C(W)$. 
\end{pr}

%In the next section it will be cleared that
%the exponent function $F(R\Vl W)$ has a close connection with
%in identification via DMC.

Arimoto \cite{ari} proved the following theorem.
\begin{Th}\ For any $R \geq 0,$\ 
${G}^{*}(R\Vl W)\geq {G}(R\Vl W)\,.$
\end{Th}

Arimoto \cite{ari} derived the lower bound ${G}(R\Vl W)$ of 
${G}^{*}(R\Vl W)$ by an analytical method. 
Subsequently, Dueck and K\"orner \cite{dk} determined  
${G}^{*}(R\Vl W)$ by a combinatorial method quite different
from that of Arimoto. To state their result for 
$P\in{\cal P}({\cal X})$ and $R\geq 0$, we define 
the following function

\beqno 
\tilde{F}_{\prmt}^{+}(R,P\Vl W)
&\defeq & 
\min_{\scs V\in {\cal P}({\cal Y}|{\cal X})}
\left\{[\prmt\left(-R+I(P;V)\right)]^{+}
\right.
\nonumber\\
&& \left. \qquad\qquad +D(V||W|P) \right\}\,,
\eeqno
where ${\cal P}({\cal Y}|{\cal X})$ is a set of all noisy channels 
with input ${\cal X}$ and output ${\cal Y}$ and $[a]^{+}=\max\{a,0\}$.
Furthermore, for $R\geq 0$, define 
\beqno 
\tilde{G}_{-1}^{+}(R\Vl W)
&\defeq& 
\min_{P\in {\cal P}({\cal X})}\tilde{F}_{-1}^{+}(R,P\Vl W)\,,
\eeqno
and for $0 \leq R \leq \log |{\cal X}|$, define
\beqno
\tilde{G}_{\mSP}(R\Vl W)&\defeq& \min_{P\in {\cal P}({\cal X})}
\min_{\scs V\in {\cal P}({\cal Y}|{\cal X}):
            \atop{\scs I(P;V)\geq R}} D(V||W|P)\,.
\eeqno
The suffix ``sp'' of the function $\tilde{G}_{\mSP}(R\Vl W)$ derives from 
that it has a form of {\it the sphere packing exponent function}. 
Those functions satisfy the following. 

\begin{pr}\label{pr:pr4}$\quad$ 
\begin{itemize}
\item[\rm{a)}] The function $\tilde{G}_{\mSP}(R\Vl W)$ is  monotone 
increasing for $0\leq R \leq \log |{\cal X}| $ and takes positive value 
if and only if $R > C(W)$.  
\item[{\rm b)}] For $0\leq R \leq \log |{\cal X}|$, we have 
$$
\tilde{G}_{-1}^{+}(R\Vl W)=\tilde{G}_{\mSP}(R\Vl W)\,.
$$
Furthermore, for $R \geq \log |{\cal X}|$, we have 
$$
\tilde{G}_{-1}^{+}(R\Vl W)=\tilde{G}_{-1}(R\Vl W)\,.
$$
\item[{\rm c)}] For $R\geq 0$
$$
|\tilde{G}_{-1}^{+}(R\Vl W)-\tilde{G}_{-1}^{+}(R^{\prime}\Vl W)|
\leq |R-R^{\prime}|\,.
$$
\end{itemize}
\end{pr}

{\it Proof: } Property \ref{pr:pr4} part a) is obvious. 
Proof of part c) 
is found in Dueck and K\"orner \cite{dk}. In this paper 
we prove the part b). To prove the first inequality, 
for fixed $P\in {\cal P}({\cal X})$, we set   
\beqno 
& &\tilde{G}_{\mSP}(R,P\Vl W)\defeq 
\min_{
\scs V\in {\cal P}({\cal Y}|{\cal X}):
            \atop{\scs I(P;V)\geq R}
}
D(V||W|P)
\\
& &\hat{F}_{-1}(R,P\Vl W)
\\
&\defeq& 
\min_{
\scs V\in {\cal P}({\cal Y}|{\cal X}):
            \atop{\scs I(P;V)\leq R}
}
\left\{R-I(P;V) +D(V||W|P) \right\}\,.
\eeqno
It is obvious that 
\beqa
\lefteqn{\tilde{F}_{-1}^{+}(R,P\Vl W)}
\nonumber\\
&=& 
\min\left\{ 
\tilde{G}_{\mSP}(R,P\Vl W), \hat{F}_{-1}(R,P\Vl W)
\right\}
\label{eqn:aa0}\\
\lefteqn{
\tilde{G}_{\mSP}(R\Vl W)= \min_{P\in {\cal P}({\cal X})}
\tilde{G}_{\mSP}(R,P\Vl W).}
\label{eqn:aa1}
\eeqa
Since  
$-I(P;V)$ 
$+D(V||W|P)$ 
is a linear function of $V$, the minimum is attained by some 
$V$ satisfying $I(P;V)=R$. Then, by (\ref{eqn:aa0}), we have
$$
\tilde{F}_{-1}^{+}(R,P\Vl W)= \tilde{G}_{\mSP}(R,P\Vl W)\,.
$$
From the above equality and (\ref{eqn:aa1}), we obtain the first equality. 
The second equality is obvious since $R-I(P;V)\geq 0$ when
$R\geq\log|{\cal X}|$.
$\quad$\hfill\QED

Dueck and K\"orner \cite{dk} proved the following. 
\begin{Th}\label{th:thDK}\ For any $R>0$,
$$
\tilde{G}_{-1}^{+}(R\Vl W)={G}^{*}(R\Vl W)\,. 
$$
\end{Th}

Although the lower bound derived by Arimoto \cite{ari} is a form quite 
different from the optimal exponent determined by Dueck and K\"orner 
\cite{dk}, the former coincides with the latter, i.e., the following 
theorem holds. 
\begin{Th}{\label{th:th1} \ For any $R\geq 0$, 
$$
\tilde{G}_{-1}^{+}(R\Vl W)={\empty}{G}(R\Vl W)\,, 
$$
or equivalent to
\beqno
& &\max_{-1 \leq \prmt \leq 0}
   \min_{P\in {\cal P}({\cal X})}
   \Hugel-\delta R
\\
& &\qquad\quad
\left. -\log
\sum_{y \in {\cal Y} }
\left[\sum_{x \in {\cal X}} 
P(x)W(y|x)^{\frac{1}{1+\prmt}} \right]^{1+\prmt}
\right\}
\\
&=&\min_{P\in {\cal P}({\cal X})}
   \min_{\scs V\in {\cal P}({\cal Y}|{\cal X})}
\ba[t]{l}
\left\{[R-I(P;V)]^{+} 
\right.
\\
\left. \qquad +D(V||W|P) \right\}\,.
\ea
\eeqno
}
\end{Th}

The result of Theorem \ref{th:th1} is stated in Csisz\'ar and K\"orner 
\cite{ck} without proof. Dueck and K\"orner \cite{dk} stated that the 
equivalence between their bound and that of Arimoto \cite{ari} can be 
proved by an analytical computation. In the next section we give a 
rigorous proof of the above theorem. Contrary to their statement, our 
proof is {\it not completely analytical}.

\section{Proof of Theorem \ref{th:th1}}

In this section we prove Theorem \ref{th:th1}.
The following is a key lemma for the proof.

\begin{lm}{\label{lm:lm0}
The function $\tilde{G}_{-1}^{+}(R\Vl W)$ is a monotone 
increasing and convex function of $R\geq 0$.
}\end{lm}

{\it Proof:} The results follows from the convexity of ${G}^{*}(R\Vl W)$ and 
Theorem \ref{th:thDK}.
\hfill\QED

\begin{rem}{\rm \ 
We first tried to prove Lemma \ref{lm:lm0} by an analytical computation 
but could not succeed proving this lemma via this approach. 
%On the convexity of $\tilde{G}_{-1}^{+}(R\Vl W)$, 
%we first tried to prove it by an analytical computation 
%but could not have succeeded in the analytical proof.
According to \cite{ck3}, for each fixed $P\in {\cal P}({\cal X})$, 
$\tilde{F}_{-1}^{+}(R,P \Vl W)$ is a convex function of $R\geq 0$. 
However, this does not imply the contexity 
of $\tilde{G}_{-1}^{+}(R\Vl W)$ with respect to $R\geq 0$. 
}
\end{rem}

Next, for $R\geq 0$, we set 
\beqno 
\tilde{F}_\prmt(R,P\Vl W)&\defeq& 
\min_{\scs V\in {\cal P}({\cal Y}|{\cal X})}
\left\{\prmt[I(P;V)-R]
\right.
\nonumber\\
&& \left. \qquad\qquad +D(V||W|P) \right\}\,,\\
\tilde{G}_\prmt(R\Vl W)&\defeq & \min_{P\in {\cal P}({\cal X})}
\tilde{F}_\prmt(R,P\Vl W)\,.
\eeqno
Then, we have the following two lemmas.
\begin{lm}\label{lm:lm4} \ 
%Assume that 
%$\tilde{G}_{-1}^{+}(R\Vl W)$ 
%is a monotone increasing and convex function of $R\geq 0$.
%Then, 
For any $R \geq 0$,
$$
\tilde{G}_{-1}^{+}(R\Vl W)=\max_{-1\leq \prmt \leq 0}
\tilde{G}_\prmt(R\Vl W)\,. 
$$
\end{lm}
\begin{lm}\label{lm:lm4c} 
For any $R\geq 0$,
$-1\leq\prmt \leq 0$ and any 
$P\in{\cal P}({\cal X})$, we have 
$$
\tilde{F}_\prmt (R,P\Vl W) \geq {\empty}{F}_{\prmt}(R,P\Vl W).
$$
Furthermore, for any $R\geq 0$ and $-1\leq\prmt\leq 0$, 
$$
\tilde{G}_{\prmt}(R\Vl W)={\empty}{G}_{\prmt}(R\Vl W)\,.
$$
\end{lm}

It is obvious that Theorem \ref{th:th1} immediately follows
from Lemmas \ref{lm:lm4} and \ref{lm:lm4c}. Those two lemmas 
can be proved by {\it analytical computations}. 
In the following we prove Lemma \ref{lm:lm4}. The proof of 
Lemma \ref{lm:lm4c} is omitted here. For the detail see 
Oohama \cite{ohID}.

{\it Proof of Lemma \ref{lm:lm4}:}
From its formula, it is obvious that
$$
\tilde{G}_{-1}^{+}(R\Vl W) 
\geq \max_{-1\leq  \prmt \leq 0} \tilde{G}_{\prmt}(R\Vl W)\,.
$$
In particular, from Property \ref{pr:pr4} part b), the equality 
holds for $R\geq \log |{\cal X}|$. Then, again by Property 
\ref{pr:pr4} part b), it suffices to prove that 
for $0\leq  R\leq \log |{\cal X}|,$ there exists 
$-1\leq  \delta\leq 0$ such that
$$
\tilde{G}_{\mSP}(R\Vl W)=\tilde{G}_\delta(R\Vl W)\,.
$$
For $-1\leq \delta \leq 0$, we set   
\beqno
& &K_{\prmt}(W)
\nonumber\\
& \defeq & \max_{P\in {\cal P}({\cal X})}
\max_{V\in {\cal P}({\cal Y}|{\cal X})} 
\left\{-\delta I(P;V)- D(V||W|P)\right\}\,.
\eeqno
Then, by the definition of $\tilde{G}_{\prmt}(R\Vl W)$, we have 
the following.
$$
\tilde{G}_{\prmt}(R\Vl W)=-\delta R - K_{\prmt}(W)\,.
$$
Next, observe that by Property \ref{pr:pr4} part b) and 
Lemma \ref{lm:lm0}, $\tilde{G}_{\mSP}(R\Vl W)$ is a monotone increasing 
and convex function of $R$. By this property and 
Property \ref{pr:pr4} part c), for any $0\leq  R\leq \log |{\cal X}|$, 
there exists $-1\leq  \delta \leq 0$ such that for any 
$0\leq R^{\prime}$ $\leq \log |{\cal X}|$, we have 
$$
\tilde{G}_{\mSP}(R^{\prime}\Vl W) \geq \tilde{G}_{\mSP}(R\Vl W)
-\delta(R^{\prime}-R)\,.
$$
Let $(P,V)\in {\cal P}({\cal X}\times{\cal Y})$
be a joint distribution that attains $\tilde{G}(R\Vl W)$.
For any $(P^{\prime},V^{\prime})\in {\cal P}({\cal X}\times{\cal Y})$
set $R^{\prime}=I(P^{\prime};V^{\prime})$. Then, we have the
following chain of inequalities:   
\beqno
& &\delta I(P^{\prime};V^{\prime})-D(V^{\prime}||W|P^{\prime}) 
\\
&\leq & -\delta R^{\prime} -\tilde{G}_{\mSP}(R^{\prime}\Vl W) 
\leq -\delta R -\tilde{G}_{\mSP}(R\Vl W) 
\\
&\leq & -\delta I(P;V)-D(V||W|P) \,.
\eeqno
The above inequality implies that 
\beqno
K_\prmt(W)&=& -\delta I(P;V) - D(V||W|P) 
\nonumber\\
 &=& - \delta R-\tilde{G}_{\mSP}(R\Vl W) \,.
\eeqno
This completes the proof. 
$\quad$\hfill\QED

\section{Comparison with the Proof of the Dual Result}
%In this section we state the result 

Theorem \ref{th:th1} has some duality with a result stated 
in Csisz\'ar and K\"orner \cite{ck}. To describe their result 
we define 
\beqno
{\empty}{E}_{\prmt}(R\Vl W)& 
\defeq &\max_{P\in {\cal P}({\cal X})}
{\empty}{F}_\prmt(R, P\Vl W)\,,
\\
{\empty}{E}(R\Vl W)
&\defeq& 
\max_{\prmt \geq 0}
{\empty}{E}_{\prmt}(R\Vl W)
\\
&=& 
\max_{\prmt \geq 0}
\max_{P\in {\cal P}({\cal X})}{\empty}{F}_{\prmt}(R,P\Vl W)
\\
&=&
\max_{\prmt \geq 0}
\left[
{-\prmt R}+\max_{P\in{\cal P}({\cal X})}
{J}_{\prmt}(P\Vl W)
\right]\,.
\eeqno
An explicit lower bound of $E^{*}(R|W)$ is first 
derived by Gallager \cite{ga2}. He showed 
that the function 
$
\max_{0\leq \prmt \leq 1}
{\empty}{E}_{\prmt}(R\Vl W)
$ 
serves as an lower bound of $E^{*}(R|W)$.
Next, we set
\beqno
C_0(W)&\defeq &\max_{P\in {\cal P}({\cal X})}
                 \min_{V\in {\cal P}({\cal Y}|{\cal X})}
I(P;V)
\eeqno
According to Shannon, Gallager and Berlekamp \cite{sgb}, $C_0(W)$ 
has the following formula:
\beqno
C_0(W)&=&-\min_{P\in{\cal P}({\cal X})}
\max_{y\in{\cal Y}}\log \sum_{x \in{\cal X}: W(y|x)>0}P(x)\,.
\eeqno
For $R\geq C_0(W)$, define
\beqno
\tilde{E}_{\SP}(R\Vl W)&\defeq & \max_{P\in {\cal P}({\cal X})}
\min_{\scs V\in {\cal P}({\cal Y}|{\cal X}):
            \atop{\scs I(P;V) \leq R}} D(V||W|P)\,.
\eeqno
According to Csisz\'ar and K\"orner \cite{ck}, 
$\tilde{E}_{\SP}(R\Vl W)$ serves as an upper bound of $E^{*}(R\Vl W)$ 
and matches it for large $R$ below $C(W)$. 
Csisz\'ar and K\"orner \cite{ck} obtained the following result.

\begin{Th}[Csisz\'ar and K\"orner \cite{ck}]
\label{th:th2}{\rm For any $R\geq C_0(W$ $)$, 
$$
{\empty}{E}(R\Vl W)=\tilde{E}_{\SP}(R\Vl W)\,,
$$
or equivalent to
\beqno
%\lefteqn{\hspace*{-12mm}
&&\max_{\prmt \geq 0}
   \max_{P\in {\cal P}({\cal X})}
   \Hugel -\delta R 
\\
& & \qquad
-\log \left.
\sum_{y \in {\cal Y} }
\left[\sum_{x \in {\cal X}} 
P(x)W(y|x)^{\frac{1}{1+\prmt}} \right]^{1+\prmt}
%\right\} 
\right\}
%}
\\
&=& \max_{P\in {\cal P}({\cal X})}
    \min_{\scs V\in {\cal P}({\cal Y}|{\cal X}):
           \atop{\scs I(P;V) \leq R}} D(V||W|P)\,.
\eeqno
}
\end{Th}

In the following we outline the arguments of the proof of the 
above theorem and compare them with those of the proof 
of Theorem \ref{th:th1}. 

By an analytical computation we have the following lemma. 
\begin{lm}{\label{lm:lmd0}
The function $\tilde{E}_{\SP}(R\Vl W)$ is a monotone 
decreasing and convex function of $R\geq C_0(W)$ and 
is positive if and only if $C_0(W) \leq  R < C(W)$. 

}\end{lm}

Next, for $R\geq 0$, we define
\beqno 
\tilde{E}_\prmt(R\Vl W)&=& \min_{P\in {\cal P}({\cal X})}
\tilde{F}_\prmt(R,P\Vl W)\,.
\eeqno
Then, we have the following two lemmas
\begin{lm}\label{lm:lmd4} \ 
%Assume that 
%$\tilde{E}_{\SP}(R\Vl W)$ is a convex function of $R\geq C_0(W).$
%Then, 
For any $R \geq C_0(W)$,
$$
\tilde{E}_{\SP}(R\Vl W)=\max_{ \prmt \geq 0}
\tilde{E}_\prmt(R\Vl W)\,. 
$$
\end{lm}
\begin{lm}\label{lm:lmd4c} 
For any $R\geq 0$, $\prmt \geq 0$ and any 
$P\in{\cal P}({\cal X})$, we have 
$$
\tilde{F}_\prmt (R,P\Vl W) \geq {\empty}{F}_{\prmt}(R,P\Vl W).
$$
Furthermore, for any $R\geq 0$ and $\prmt\geq 0$, 
$$
\tilde{E}_{\prmt}(R\Vl W)={\empty}{E}_{\prmt}(R\Vl W)\,.
$$
\end{lm}

It is obvious that Theorem \ref{th:th2} immediately follows
from Lemmas \ref{lm:lmd4} and \ref{lm:lmd4c}. 
We prove Lemmas \ref{lm:lmd4} and \ref{lm:lmd4c} 
in manners quite similar to those of the proofs 
of Lemmas \ref{lm:lm4} and \ref{lm:lm4c}, respectively. 
We omit the details of the proofs. 

We compare the arguments of the proof of Theorem \ref{th:th1} 
with those of the proof of Theorem \ref{th:th2}. An essential 
difference between them is in the proof of the convexity of
exponent functions. We can prove the convexity of 
$\tilde{E}_{\SP}(R\Vl W)$ with an analytical method. On 
the other hand, the convexity $\tilde{G}_{-1}^{+}(R\Vl W)$ 
follows from ${G}^{*}(R\Vl W)=$$\tilde{G}_{-1}^{+}(R\Vl W)$
and the convexity of ${G}^{*}(R\Vl W)$. 
The proof of the convexity of ${G}^{*}(R\Vl W)$ is based on 
an operational meaning of the optimal exponent function
of $1-\varepsilon_n$. We first tried an analytical proof of 
the convexity $\tilde{G}_{-1}^{+}(R\Vl W)$ but could not 
have succeeded in it. The difference of arguments is 
summarized in TABLE \ref{tab:tb1}.

 \renewcommand{\irg}[1]{{\color[named]{Black}#1\normalcolor}}
 \renewcommand{\irb}[1]{{\color[named]{Black}#1\normalcolor}}
\renewcommand{\irBr}[1]{{\color[named]{Black}#1\normalcolor}}

\begin{table}[t]
\newcommand{\lw}[1]{\smash{\lower2.ex\hbox{#1}}}
\renewcommand{\arraystretch}{1.4}
\doublerulesep 0.3mm
\tabcolsep 1.4mm
\begin{tabular}{|c|c|}
\hline
$\irb{R}>\irb{C(\irBr{W})}$ & $\irb{R}<\irb{C(\irBr{W})}$
\\ \hline\hline
{\renewcommand{\arraystretch}{1.5}
\begin{tabular}{c}
$G^{*}(\irb{R}\Vl \irBr{W})=\tilde{G}_{-1}^{+}(\irb{R}\Vl \irBr{W})$
\vspace{-1mm}\\
(\irb{Theorem \ref{th:thDK}})
\end{tabular}}
& {\renewcommand{\arraystretch}{1.5}
\begin{tabular}{c}
$E^{*}(\irb{R}\Vl \irBr{W}) \leq \tilde{E}_{\SP}(R\Vl W)$
\vspace{-1mm}\\ 
(Open Problem )
\end{tabular}}
\\ \hline
\irg{Operational Meaning} & 
\\
$\Downarrow$ & \lw{Convexity of $E^{*}(\irb{R}\Vl \irBr{W})$ ?}\\ 
{\renewcommand{\arraystretch}{1.2}
\begin{tabular}{c}
Convexity of ${G}^{*}(\irb{R}\Vl \irBr{W})$ \\
(Property \ref{pr:pr0}) 
\end{tabular}}
& 
\\ \hline
\irb{Theorem \ref{th:thDK}} and Property \ref{pr:pr0} & 
\irg{Analytical Computation}
\\
$\Downarrow$ & $\Downarrow$
\\
  {\renewcommand{\arraystretch}{1.1}
  \begin{tabular}{c}
    Convexity of $\tilde{G}_{-1}^{+}(\irb{R}\Vl \irBr{W})$\\
    (Lemma \ref{lm:lm0}) 
  \end{tabular}}
 &  {\renewcommand{\arraystretch}{1.1}
    \begin{tabular}{c}
    Convexity of $\tilde{E}_{\SP}(\irb{R}\Vl \irBr{W})$\\
    (Lemma \ref{lm:lmd0})
    \end{tabular}}
\\  %\hline
$\Downarrow$ & $\Downarrow$
\\
  \irb{Lemma \ref{lm:lm4}} 
& \irb{Lemma \ref{lm:lmd4}} 
\\ \hline
  \irb{Lemmas \ref{lm:lm4}  and \ref{lm:lm4c}} 
& \irb{Lemmas \ref{lm:lmd4} and \ref{lm:lmd4c}} 
\\
$\Downarrow$ & $\Downarrow$\\
   {\renewcommand{\arraystretch}{1.1}
   \begin{tabular}{c}
   $G(\irb{R}\Vl \irBr{W})=\tilde{G}_{-1}^{+}(\irb{R}\Vl \irBr{W})$ \\
   (Theorem \ref{th:th1})
   \end{tabular}}
&  {\renewcommand{\arraystretch}{1.1}
   \begin{tabular}{c}
   $E(\irb{R}\Vl \irBr{W})=\tilde{E}_{\SP}(\irb{R}\Vl \irBr{W})$ \\
   (Theorem \ref{th:th2} )
   \end{tabular}}
\\ \hline
\end{tabular}

\caption{
Difference between the arguments of the proof of Theorem \ref{th:th1} 
and those of the proof of Theorem \ref{th:th2}. 
}
\label{tab:tb1}
%}
%\end{flushleft}  
\end{table}

\end{document}